\documentclass[twocolumn,amsmath,amssymb]{revtex4}
\usepackage{graphicx}
\usepackage{dcolumn}
\usepackage{bm}

\begin{document}


\title{Posterior probability and fluctuation theorem in stochastic processes}

\author{Jun Ohkubo} 
\email[Email address: ]{ohkubo@issp.u-tokyo.ac.jp}
\affiliation{
Institute for Solid State Physics, University of Tokyo, 
Kashiwanoha 5-1-5, Kashiwa-shi, Chiba 277-8581, Japan
}

\date{\today}

\begin{abstract}
A generalization of fluctuation theorems in stochastic processes is proposed.
The new theorem is written in terms of posterior probabilities,
which are introduced via the Bayes theorem.
In usual fluctuation theorems, 
a forward path and its time reversal play an important role,
so that a microscopically reversible condition is essential.
In contrast, the microscopically reversible condition is not necessary in the new theorem.
It is shown that the new theorem adequately recovers 
various theorems and relations previously known,
such as the Gallavotti-Cohen-type fluctuation theorem, the Jarzynski equality, and the Hatano-Sasa relation,
when adequate assumptions are employed.
\end{abstract}

\pacs{}

\keywords{Bayes theorem, posterior probability, fluctuation theorem, Gallavotti-Cohen type symmetry,
stochastic process, nonequilibrium systems}

\maketitle

The discovery of fluctuation theorems is one of the most important developments 
in nonequilibrium physics.
(There are many review articles; for example, see refs.~1 and 2 and references therein.)
The fluctuation theorems are applicable not only in near equilibrium but also far from equilibrium,
and hence it is believed that the fluctuation theorems play an important role
in the study for nonequilibrium physics.
There are some types of fluctuation theorems,
such as a Gallavotti-Cohen (GC) fluctuation theorem 
\cite{Gallavotti1995,Gallavotti1995a,Kurchan1998,Lebowitz1999,Gaspard2004}
and an integral fluctuation theorem \cite{Crooks1998,Crooks1999,Crooks2000,Seifert2005}.
The GC fluctuation theorem is related to a symmetry of a certain quantity,
and applicable if one considers the average of the quantity over a large time span.
In contrast, the integral fluctuation theorem can be used for any finite time span.
It has been known that the integral fluctuation theorem directly gives 
the Jarzynski equality, which connects a free energy difference
between two equilibrium states and a non-equilibrium work required to move
a equilibrium state to the other one in a finite time
\cite{Jarzynski1997,Jarzynski1997a,Crooks1998,Crooks1999,Crooks2000}.

Usually, the fluctuation theorems focus on
a quantity which is defined by a forward path and its time reversal \cite{Crooks1998},
and the average of the quantity over all forward paths are taken.
The existence of the time reversal means 
that the system has a microscopically reversible condition \cite{Crooks1998,Crooks1999}.
However, it is an open question whether
we can obtain a kind of fluctuation theorems
when the microscopically reversible condition is not satisfied.
In addition, it has been known that the Jarzynski-like identity also holds
even in a nonequilibrium steady state;
the Hatano-Sasa relation connects a Shannon entropy difference and an excess heat
for Langevin dynamics describing nonequilibrium steady states \cite{Hatano2001}.
However, as pointed out in ref.~15, 
the integral fluctuation theorem is not directly related to the Hatano-Sasa relation.

In the present paper,
we give a new fluctuation theorem with the GC-type symmetry for stochastic processes
in terms of `posterior probabilities'.
The posterior probability is a conditional probability for an initial condition (or state),
that is assigned after an outcome (or final state) is taken into account.
The posterior probabilities always exist for arbitrary stochastic processes,
and hence the microscopically reversible condition is not needed.
The key point in the new formulation is the usage of the Bayes theorem,
which naturally introduces the posterior probabilities.
The Bayes theorem has widely been used in applications of spin glass theory 
to information processing\cite{Nishimori2001}.
In addition, there is also a work in which the Bayes theorem was used to improve 
an experimental estimation of free energy differences via the Jarzynski equality\cite{Maragakis2008}.
Another usage of the Bayes theorem in nonequilibrium physics will be shown in the present paper;
the Bayes theorem is used to avoid the usage of the microscopically reversible condition.

We will show that the new fluctuation theorem recovers various known results using adequate assumptions.
If a local detailed balance condition is employed,
the microscopically reversible condition is satisfied,
and hence the posterior probabilities are rewritten in terms of usual conditional probabilities
for time reversals.
As a consequence, the usual GC-type fluctuation theorem is recovered.
When transitions between isothermal equilibrium states are considered,
we obtain the Jarzynski equality from the new fluctuation theorem.
In addition, 
if we consider a Langevin dynamics describing nonequilibrium steady states,
the new fluctuation theorem leads to the Hatano-Sasa relation directly.

We first explain general settings and notations.
In the present paper,
a general stochastic process on a discrete set of states, $\{n\}$,
is considered.
Instead of the discrete states, it is also possible to use a continuous state space.
Extensions to continuous cases are straightforward;
summations in the following discussions
should be replaced by integrals for the continuous state space.

Assume that the stochastic system is specified by a set of control parameters, $\alpha$,
and that the system has a steady state for each $\alpha$.
The steady state probability distribution for state $n$ in the control parameter $\alpha$
is denoted as $p^\mathrm{ss}(n; \alpha)$.
Note that the steady state may be in equilibrium or in nonequilibrium, which depends on a problem
to be considered.
The system is manipulated by changing the value of $\alpha$ during the interval from $t = 0$ to $t = T$.
In addition, the system is assumed to be in a steady state initially.

According to a usual path integral formulation,
we discretize the time $t \in [0,T]$ as $t_i \in \{t_0, t_1, \dots, t_N\}$,
taking the limit $N \to \infty$ keeping $T$ fixed finally.
The value of $\alpha$ at time $t_i$ is denoted as $\alpha_i$,
and a state $n$ at time $t_i$ is written as $n_i$.
In addition, a path, which is a set of states $(n_0, n_1, \dots, n_N)$, 
is denoted as $[n]$.

A probability $\pi (n_{i+1}|n_{i} ; \alpha_{i})$ is defined
as the transition probability from $n_i$ to $n_{i+1}$ under the control parameter $\alpha_i$
in one time step.
Hence, the average of a quantity $g([n])$, which depends on a path $[n]$, is evaluated as
\begin{align}
\langle g \rangle
\simeq \sum_{\{n_i\}} \left( \prod_{i=0}^{N-1} \pi(n_{i+1}|n_{i} ; \alpha_{i}) \right)
p^\mathrm{ss} (n_{0}; \alpha_{0} )
g([n]).
\end{align}
Since the above equation would become exact after taking the limit of $N \to \infty$, 
we use the symbol $\simeq$.

In order to give the new fluctuation theorem,
new quantities $\Delta s^\mathrm{sys}$, $\Delta s^\mathrm{m}$, and 
$\Delta s^\mathrm{tot}$ are introduced as follows:
\begin{align}
&\Delta s^\mathrm{sys} \equiv \ln p^\mathrm{ss} (n_0; \alpha_0) - \ln p(n_N; \alpha_N) ,
\label{eq_def_for_s_sys} \\
&\Delta s^\mathrm{m} \equiv \sum_{i=0}^{N-1} 
\left( \ln p^\mathrm{ss}(n_{i+1}; \alpha_{i}) - \ln p^\mathrm{ss} (n_{i}; \alpha_{i})
\right),
\label{eq_def_for_s_m} \\
&\Delta s^\mathrm{tot} \equiv \Delta s^\mathrm{sys} + \Delta s^\mathrm{m}.
\label{eq_def_for_s_total}
\end{align}
The quantity $\Delta s^\mathrm{sys}$ is a Shannon entropy difference
between the initial state and the final one.
Note that there is no need to set the final state as a steady state;
the probability distribution at final time is arbitrary,
and usually it would be determined by the time development of the system.

The main theorem in the present paper is as follows;
\begin{align}
&\sum_{\{n_i\}} \left( \prod_{i=0}^{N-1} \pi(n_{i+1}|n_{i} ; \alpha_{i}) \right)
p^\mathrm{ss} (n_{0}; \alpha_{0} ) 
\exp\left( - \lambda \Delta s^\mathrm{tot} \right) \notag \\
=&
\sum_{\{n_i\}} \left( \prod_{i=0}^{N-1} \pi^\mathrm{p} (n_{i}|n_{i+1} ; \alpha_{i}) \right)
p (n_{N}; \alpha_{N} ) \notag \\
& \times \exp\left( - (\lambda-1) \Delta s^\mathrm{tot} \right),
\label{eq_new_FT}
\end{align}
where $\pi^\mathrm{p} (n_{i}|n_{i+1} ; \alpha_{i})$ is a posterior probability.
The posterior probability is a conditional probability for a (previous) state $n_i$ 
assuming that the next state is $n_{i+1}$.
Note that the posterior probability is well defined
even if we do not have the microscopically reversible condition.
Symbolically, eq.~\eqref{eq_new_FT} is rewritten as
\begin{align}
\left\langle
\exp\left( -\lambda \Delta s^\mathrm{tot} \right)
\right\rangle
=
\left\langle
\exp\left( - (\lambda-1) \Delta s^\mathrm{tot} \right)
\right\rangle^\mathrm{p},
\label{eq_new_FT_symbol}
\end{align}
where $\langle \cdot \rangle^\mathrm{p}$
is the expectation value evaluated using the posterior probabilities.

We here give a proof of the new fluctuation theorem.
Using the Bayes theorem, the posterior probability is evaluated by
\begin{align}
\pi^\mathrm{p}(n_{i} | n_{i+1} ; \alpha_{i}) 
= 
\frac
{\pi (n_{i+1}| n_{i} ; \alpha_{i}) p^\mathrm{ss} (n_{i} ; \alpha_{i})}
{p^\mathrm{ss}(n_{i+1}; \alpha_{i})}.
\label{eq_Bayes}
\end{align}
Although the Bayes theorem is applicable if the state is not a steady one,
the posterior probability does not change depending on 
whether the prior probability is the steady one or not.
Hence, we here considered the probability in the steady state.
Replacing the conditional probability $\pi (n_{i+1}| n_{i} ; \alpha_{i})$
in the l.h.s. of eq.~\eqref{eq_new_FT} as
$\pi^\mathrm{p}(n_{i}|n_{i+1};\alpha_{i}) p^\mathrm{ss}(n_{i+1};\alpha_{i})/p^\mathrm{ss} (n_{i};\alpha_{i})$,
the relation \eqref{eq_new_FT} is immediately obtained.

Equations~\eqref{eq_new_FT} (or \eqref{eq_new_FT_symbol}) is the main result of this paper.
The microscopically reversible condition is not necessary for the new formulation,
and we therefore expect that the new fluctuation theorem~\eqref{eq_new_FT}
is an extension of usual fluctuation theorems.
Henceforth, we confirm that the new fluctuation theorem~\eqref{eq_new_FT} 
is consistent with previously known results;
we discuss connections with some of known results,
i.e., the usual GC-type fluctuation theorem, the Jarzynski equality,
and the Hatano-Sasa relation.

Firstly, we show how 
the usual GC-type fluctuation theorem is recovered from the new fluctuation theorem \eqref{eq_new_FT}.
In order to discuss the GC-type fluctuation theorem, 
we consider a situation in which the control parameter $\alpha$ remains fixed,
so that we omit $\alpha$ here.
In addition, due to the time-independent property of the system,
the system is also in a steady state at the final time step.
Employing a local detailed balance condition, 
a time reversal of $[n]$ is naturally introduced.
Because we consider a Markov process,
the forward path $[n]$ has a path probability
\begin{align}
&\mathrm{Prob}(n_0 \to n_1 \to n_2 \cdots \to n_N) \notag \\
&= \pi(n_1 | n_0) \pi(n_2 | n_1) \cdots \pi(n_N | n_{N-1}).
\end{align}
The corresponding path probability for its time reversal 
$[n]^\mathrm{rev} = (\tilde{n}_0, \tilde{n}_1, \dots \tilde{n}_N)$ is
\begin{align}
&\mathrm{Prob}(\tilde{n}_0 \to \tilde{n}_1 \to  \tilde{n}_2 \cdots \to \tilde{n}_N) \notag \\
&\equiv \mathrm{Prob}(n_0 \leftarrow n_1 \leftarrow n_2 \cdots \leftarrow n_N) \notag \\
&= \pi(n_0 | n_1) \pi(n_1 | n_2) \cdots \pi(n_{N-1} | n_{N}),
\end{align}
where $\tilde{n}_i = n_{N-i}$.
Using the transition probability for the time reversal,
the local detailed balance condition is expressed by
\begin{align}
\pi(n_{i+1}|n_{i}) p^\mathrm{ss}(n_{i})
= \pi(n_{i} | n_{i+1}) p^\mathrm{ss}(n_{i+1}),
\label{eq_LDB}
\end{align}
where $\pi(n_{i} | n_{i+1})$ is not the posterior probability,
but the usual transition probability for the time reversal.
Hence, we obtain
\begin{align}
\frac{p^\mathrm{ss}(n_{i})}{p^\mathrm{ss}(n_{i+1})}
=
\frac{\pi(n_{i} | n_{i+1})}{\pi(n_{i+1}|n_{i})}
\label{eq_rev_detailed_balance}
\end{align}
and then the quantity $\Delta s^\mathrm{m}$ is written as
\begin{align}
\Delta s^\mathrm{m}
= \ln
\left[ 
\prod_{i=0}^{N-1} \frac{\pi(n_{i+1}|n_{i})}{\pi(n_{i} | n_{i+1})}
\right].
\label{eq_def_s_m_for_GC}
\end{align}
Thus, the corresponding quantities of the time reversal
for $\Delta s^\mathrm{m}$ and $\Delta s^\mathrm{sys}$ are
\begin{align}
\Delta s^\mathrm{m,rev}
\equiv \ln
\left[ 
\prod_{i=0}^{N-1} \frac{\pi(\tilde{n}_{i+1}| \tilde{n}_{i})}
{\pi(\tilde{n}_{i} | \tilde{n}_{i+1})}
\right]
= - \Delta s^\mathrm{m}
\label{eq_def_s_m_rev}
\end{align}
and
\begin{align}
\Delta s^\mathrm{sys,rev} \equiv \ln p^\mathrm{ss}(\tilde{n}_0) - \ln^\mathrm{ss} p(\tilde{n}_N) 
= - \Delta s^\mathrm{sys}.
\label{eq_def_s_sys_rev}
\end{align}
In addition, combining the local detailed balance condition \eqref{eq_LDB}
and the Bayes theorem \eqref{eq_Bayes},
we have
\begin{align}
\pi^\mathrm{p}(n_{i}| n_{i+1}) = \pi (n_{i}| n_{i+1}).
\label{eq_posterior_and_transition}
\end{align}
Using the above identities 
(eqs.~\eqref{eq_def_s_m_rev}, \eqref{eq_def_s_sys_rev}, \eqref{eq_posterior_and_transition}),
the r.h.s of eq.~\eqref{eq_new_FT} is rewritten as
\begin{align}
&\sum_{\{n_i\}} \left( \prod_{i=0}^{N-1} \pi^\mathrm{p} (n_{i}|n_{i+1}) \right)
p^\mathrm{ss} (n_{N}) 
 \exp\left( - (\lambda-1) \Delta s^\mathrm{tot} \right) \notag \\
=&
\sum_{\{n_i\}} \left( \prod_{i=0}^{N-1} \pi(n_{i}|n_{i+1}) \right)
p^\mathrm{ss} (n_{N} ) 
\exp\left( - (\lambda-1) \Delta s^\mathrm{tot} \right)  \notag \\
=&
\sum_{\{n_i\}^\mathrm{rev}} \left( \prod_{i=0}^{N-1} \pi(n_{i+1}|n_{i}) \right)
p^\mathrm{ss} (n_{0}) \notag \\
&\times \exp\left( - (\lambda-1) \Delta s^\mathrm{sys,rev}
- (\lambda-1) \Delta s^\mathrm{m,rev} \right)  \notag \\
=&
\sum_{\{n_i\}} \left( \prod_{i=0}^{N-1} \pi(n_{i+1}|n_{i}) \right)
p^\mathrm{ss} (n_{0}) \notag \\
&\times \exp\left( - (1-\lambda) \Delta s^\mathrm{sys} 
- (1-\lambda) \Delta s^\mathrm{m} \right), 
\end{align}
where the second equality uses the fact that summing over 
the time reversal $[n]^\mathrm{rev}$ is the same as summing over the forward paths $[n]$
since both sums cover all the possible paths.
Note that,
for the third equality,
the summation for $\{n_i\}$ is equal to that of $\{n_i\}^\mathrm{rev}$.
If the final time $T$ is large enough,
one may expect that $\Delta s^\mathrm{sys}$ is negligible compared with $\Delta s^\mathrm{m}$.
If so, we obtain
\begin{align}
&
\lim_{T \to \infty} \left( 
\frac{1}{T} \ln \left\langle 
\exp\left( -\lambda \Delta s^\mathrm{m} \right)
\right\rangle\right) \notag \\
&=
\lim_{T \to \infty} \left( 
\frac{1}{T} \ln \left\langle 
\exp\left( -(1-\lambda) \Delta s^\mathrm{m} \right)
\right\rangle\right),
\end{align}
which is the GC-type fluctuation theorem given by Lebowitz and Spohn\cite{Lebowitz1999}.

Secondly, we recover the Jarzynski equality from the new fluctuation theorem~\eqref{eq_new_FT}.
In order to discuss the Jarzynski equality,
we should specify a stochastic model.
If not, thermodynamic quantities are not well-defined.
Here, we mainly obey the Crook's discussions\cite{Crooks2000}.
Assume that a system with a constant temperature heat bath.
Using the inverse temperature $\beta \equiv 1 / (k_\mathrm{B} T)$,
where $k_\mathrm{B}$ is the Boltzmann constant,
the probability for state $n_i$ under a set of control parameters $\alpha_i$
is 
\begin{align}
p^\mathrm{ss}(n_i;\alpha_i)
= \frac{\exp(-\beta E(n_{i}; \alpha_{i}))}{Z(\alpha_i)},
\end{align}
where $Z(\alpha_i)$ is the partition function.
At each time step,
the system may make a stochastic transition from state $n_i$ to $n_{i+1}$,
which has an energy difference $E(n_{i+1}; \alpha_i) - E(n_i ; \alpha_i)$.
This causes an energy transfer from the heat bath in the form of heat.
Hence, the total heat exchanged with the heat bath in a stochastic path $[n]$ is 
\begin{align}
\mathcal{Q}[n] = \sum_{i=0}^{N-1}
\left\{
E(n_{i+1}; \alpha_{i}) - E(n_{i}; \alpha_{i})
\right\}.
\end{align}
Defining the energy difference between the initial state and the final one
as 
\begin{align}
\Delta E = E(n_N; \alpha_{N}) - E(n_0; \alpha_0),
\end{align}
and the Helmholtz free energy difference as
\begin{align}
\Delta F = - \frac{1}{\beta} \ln Z(\alpha_N) + \frac{1}{\beta}  \ln Z(\alpha_0),
\end{align}
the quantity $\Delta s^\mathrm{tot}$ becomes
\begin{align}
\Delta s^\mathrm{tot} = \beta \left( 
\Delta E - \mathcal{Q} - \Delta F
\right).
\end{align}
Using the usual definition of the work performed on the system,
\begin{align}
\mathcal{W} = \Delta E - \mathcal{Q},
\end{align}
and setting $\lambda = 1$ in the new fluctuation theorem~\eqref{eq_new_FT},
we have the Jarzynski equality
\begin{align}
\left\langle \exp\left( - \beta \mathcal{W} \right) \right\rangle
= \exp \left( -\beta \Delta F \right).
\end{align}

Thirdly, we discuss a connection with the Hatano-Sasa relation.
We consider a Langevin dynamics driven by an external force
\begin{align}
\gamma \frac{\mathrm{d}}{\mathrm{d} t} x
= - \frac{\partial U(x;\theta)}{\partial x} + f + \xi(t),
\end{align}
where $\xi(t)$ represent Gaussian white noise whose intensity is $2\gamma k_\mathrm{B} T$.
The external force $f$ and a control parameter for potential, $\theta$,
specifies the nonequilibrium steady state of the Langevin dynamics;
we define a set of control parameters as $\alpha = (f,\theta)$.
In the nonequilibrium case, it is impossible to write the probability in the steady state
in the form of the Boltzmann distribution.
Hence, we merely write it as $\rho_\mathrm{ss}(x; \alpha)$.
As explained before, the new fluctuation theorem \eqref{eq_new_FT_symbol} is also applicable 
to the continuous state space.
Setting $\lambda = 1$ and the final state as a steady state,
we have
\begin{align}
\left\langle
\prod_{i=0}^{N-1} \frac{\rho_\mathrm{ss}(x_{i+1};\alpha_{i+1})}{\rho_\mathrm{ss}(x_{i+1};\alpha_i)}
\right\rangle
\simeq 1,
\end{align}
which is actually the starting point for the derivation of the Hatano-Sasa relation;
see eq.~(9) in ref.~15.
Thus, the new fluctuation theorem immediately recovers the Hatano-Sasa relation.
The Hatano-Sasa relation is expressed as follows:
\begin{align}
\left\langle
\exp\left[
-\beta \mathcal{Q}_\mathrm{ex} - \Delta \phi
\right]
\right\rangle = 1
\end{align}
$\mathcal{Q}_\mathrm{ex}$ is the excess heat\cite{Oono1998,Hatano2001}
which is defined by subtracting the housekeeping heat from the total heat,
and $\Delta \phi$ is the Shannon entropy difference 
between the initial and final states\cite{Hatano2001}.
Although the usual fluctuation theorem is not directly related
to the Hatano-Sasa relation, as denoted in ref.~15,
the new fluctuation theorem~\eqref{eq_new_FT} adequately gives the Hatano-Sasa relation.

In conclusion,
we showed a new fluctuation theorem in terms of posterior probabilities.
The microscopically reversible condition,
which is used in usual fluctuation theorems,
is not necessary for the new fluctuation theorem.
We confirmed that the new fluctuation theorem adequately recovers
previously known results,
i.e., the GC-type fluctuation theorem, Jarzynski equality, and Hatano-Sasa relation
when we consider adequate stochastic models, respectively.
We expect that this new fluctuation theorem would be applicable
for some chemical reactions without the microscopically reversible condition.
Whenever each chemical reaction does not have its reverse reaction,
the usual fluctuation theorems cannot be used
because it is impossible to define a time reversal.
The new fluctuation theorem is available even in such cases;
the applications of the new fluctuation theorem are future works.

Finally, we comment that
the usage of Bayes theorem and the posterior probabilities
may open up new possibilities for the study of nonequilibrium physics.
Actually, 
the posterior probabilities have also been used
in the study of nonequilibrium steady states\cite{Komatsu2008,Komatsu2008a,Komatsu2009}.
As explained, the posterior probabilities are naturally introduced
via the Bayes theorem, without the microscopically reversible condition.
Although it is an open problem whether the usage of the posterior probabilities is crucial 
in nonequilibrium physics or not,
at least we clarified that many theorems and relationships 
are recovered from the new fluctuation theorem in a unified way.
In addition,
it may be possible to determine (forward) conditional probabilities experimentally;
performing many trials and measuring initial and final states,
the conditional probabilities would be evaluated.
Similarly, the posterior probabilities may also be measurable.
If one considers a specific stochastic model,
the posterior probabilities may be calculated analytically.
However, as pointed out in ref.~19, the evaluation of the posterior probabilities
will be difficult practically, even in numerical calculations;
it could be an interesting problem to evaluate the posterior probabilities efficiently.

We hope that the new formulation
gives further insights and techniques for studies of nonequilibrium physics.

\section*{Acknowledgment}
The author is grateful to S. Sasa for giving useful comments in the first stage of this work.
This work was supported in part by grant-in-aid for scientific research 
(Grants No.~20115009 and No.~21740283)
from the Ministry of Education, Culture, Sports, Science and Technology, Japan.

\end{document}